\documentstyle[art11]{article}
\advance \topmargin by -\headheight
\advance \topmargin by -\headsep

\evensidemargin \oddsidemargin
\begin{document}



\def\rf#1{(\ref{eq:#1})}
\def\lab#1{\label{eq:#1}}
\def\nonu{\nonumber}
\def\br{\begin{eqnarray}}
\def\er{\end{eqnarray}}
\def\be{\begin{equation}}
\def\ee{\end{equation}}
\def\eq{\!\!\!\! &=& \!\!\!\! }
\def\foot#1{\footnotemark\footnotetext{#1}}
\def\lb{\lbrack}
\def\rb{\rbrack}
\def\llangle{\left\langle}
\def\rrangle{\right\rangle}
\def\blangle{\Bigl\langle}
\def\brangle{\Bigr\rangle}
\def\llb{\left\lbrack}
\def\rrb{\right\rbrack}
\def\Blb{\Bigl\lbrack}
\def\Brb{\Bigr\rbrack}
\def\lcurl{\left\{}
\def\rcurl{\right\}}
\def\({\left(}
\def\){\right)}
\def\v{\vert}                     
\def\bv{\bigm\vert}               
\def\Bgv{\;\Bigg\vert}            
\def\bgv{\bigg\vert}              
\def\lskip{\vskip\baselineskip\vskip-\parskip\noindent}
\def\mskp{\par\vskip 0.3cm \par\noindent}
\def\sskp{\par\vskip 0.15cm \par\noindent}
\def\bc{\begin{center}}
\def\ec{\end{center}}
\def\Lbf#1{{\Large {\bf {#1}}}}
\def\lbf#1{{\large {\bf {#1}}}}


\def\tr{\mathop{\rm tr}}                  
\def\Tr{\mathop{\rm Tr}}                  
\newcommand\partder[2]{{{\partial {#1}}\over{\partial {#2}}}}
\newcommand\partderd[2]{{{\partial^2 {#1}}\over{{\partial {#2}}^2}}}
\newcommand\partderh[3]{{{\partial^{#3} {#1}}\over{{\partial {#2}}^{#3} }}}
\newcommand\partderm[3]{{{\partial^2 {#1}}\over{\partial {#2} \partial {#3} }}}
\newcommand\partderM[6]{{{\partial^{#2} {#1}}\over{{\partial {#3}}^{#4}
{\partial {#5}}^{#6} }}}          
\newcommand\funcder[2]{{{\delta {#1}}\over{\delta {#2}}}}
\newcommand\Bil[2]{\Bigl\langle {#1} \Bigg\vert {#2} \Bigr\rangle}  
\newcommand\bil[2]{\left\langle {#1} \bigg\vert {#2} \right\rangle} 
\newcommand\me[2]{\left\langle {#1}\right|\left. {#2} \right\rangle} 

\newcommand\sbr[2]{\left\lbrack\,{#1}\, ,\,{#2}\,\right\rbrack} 
\newcommand\Sbr[2]{\Bigl\lbrack\,{#1}\, ,\,{#2}\,\Bigr\rbrack} 
\newcommand\pbr[2]{\{\,{#1}\, ,\,{#2}\,\}}       
\newcommand\Pbr[2]{\Bigl\{ \,{#1}\, ,\,{#2}\,\Bigr\}}  
\newcommand\pbbr[2]{\lcurl\,{#1}\, ,\,{#2}\,\rcurl}  


\def\a{\alpha}
\def\b{\beta}
\def\c{\chi}
\def\d{\delta}
\def\D{\Delta}
\def\eps{\epsilon}
\def\vareps{\varepsilon}
\def\g{\gamma}
\def\G{\Gamma}
\def\grad{\nabla}
\def\h{{1\over 2}}
\def\l{\lambda}
\def\L{\Lambda}
\def\m{\mu}
\def\n{\nu}
\def\ov{\over}
\def\om{\omega}
\def\O{\Omega}
\def\p{\phi}
\def\P{\Phi}
\def\pa{\partial}
\def\tpa{{\tilde \partial}}
\def\pr{\prime}
\def\ra{\rightarrow}
\def\lra{\longrightarrow}
\def\s{\sigma}
\def\S{\Sigma}
\def\t{\tau}
\def\th{\theta}
\def\Th{\Theta}
\def\z{\zeta}
\def\ti{\tilde}
\def\wti{\widetilde}
\newcommand\sumi[1]{\sum_{#1}^{\infty}}   
\newcommand\fourmat[4]{\left(\begin{array}{cc}  
{#1} & {#2} \\ {#3} & {#4} \end{array} \right)}
\newcommand\twocol[2]{\left(\begin{array}{cc}  
{#1} \\ {#2} \end{array} \right)}


\def\phanta{\phantom{aaaaaaaaaaaaaaa}}
\def\phantb{\phantom{aaaaaaaaaaaaaaaaaaaaaaaaa}}
\def\phantc{\phantom{aaaaaaaaaaaaaaaaaaaaaaaaaaaaaaaaaaa}}


\font\numbers=cmss12
\font\upright=cmu10 scaled\magstep1
\def\stroke{\vrule height8pt width0.4pt depth-0.1pt}
\def\topfleck{\vrule height8pt width0.5pt depth-5.9pt}
\def\botfleck{\vrule height2pt width0.5pt depth0.1pt}
\def\Zmath{\vcenter{\hbox{\numbers\rlap{\rlap{Z}\kern 0.8pt\topfleck}\kern
2.2pt
                   \rlap Z\kern 6pt\botfleck\kern 1pt}}}
\def\Qmath{\vcenter{\hbox{\upright\rlap{\rlap{Q}\kern
                   3.8pt\stroke}\phantom{Q}}}}
\def\Nmath{\vcenter{\hbox{\upright\rlap{I}\kern 1.7pt N}}}
\def\Cmath{\vcenter{\hbox{\upright\rlap{\rlap{C}\kern
                   3.8pt\stroke}\phantom{C}}}}
\def\Rmath{\vcenter{\hbox{\upright\rlap{I}\kern 1.7pt R}}}
\def\IZ{\ifmmode\Zmath\else$\Zmath$\fi}
\def\IQ{\ifmmode\Qmath\else$\Qmath$\fi}
\def\IN{\ifmmode\Nmath\else$\Nmath$\fi}
\def\IC{\ifmmode\Cmath\else$\Cmath$\fi}
\def\IR{\ifmmode\Rmath\else$\Rmath$\fi}

\def\one{\hbox{{1}\kern-.25em\hbox{l}}}
\def\0#1{\relax\ifmmode\mathaccent"7017{#1}%
        \else\accent23#1\relax\fi}
\def\omz{\0 \omega}


\def\mark{\noindent{\bf Remark.}\quad}
\def\prop{\noindent{\bf Proposition.}\quad}
\def\exam{\noindent{\bf Example.}\quad}

\newtheorem{definition}{Definition}[section]
\newtheorem{proposition}{Proposition}[section]
\newtheorem{theorem}{Theorem}[section]
\newtheorem{lemma}{Lemma}[section]
\newtheorem{corollary}{Corollary}[section]
\def\proof{\par{\it Proof}. \ignorespaces} \def\endproof{{$\Box$}\par}
\newenvironment{Proof}{\proof}{\endproof}


\def\Winf{{\bf W_\infty}}               
\def\Win1{{\bf W_{1+\infty}}}           
\def\nWinf{{\bf {\hat W}_\infty}}       
\def\hWinf{{\bf {\hat W}_{\infty}}}        


\def\PsDA{\Psi{\cal DO}}

\def\cKP{{\sf cKP}~}
\def\scKP{{\sf scKP}~}
\newcommand\Back{{B\"{a}cklund}~}
\newcommand\DB{{Darboux-B\"{a}cklund}~}
\def\BH{{Burgers-Hopf}~}
\def\tQ{{\widetilde Q}}
\def\tit{{\tilde t}}
\def\hQ{{\widehat Q}}
\def\hb{{\widehat b}}
\def\hR{{\widehat R}}
\def\htt{{\hat t}}

\newcommand{\nit}{\noindent}
\newcommand{\ct}[1]{\cite{#1}}
\newcommand{\bi}[1]{\bibitem{#1}}
%
%
\newcommand\PRL[3]{{\sl Phys. Rev. Lett.} {\bf#1} (#2) #3}
\newcommand\NPB[3]{{\sl Nucl. Phys.} {\bf B#1} (#2) #3}
\newcommand\NPBFS[4]{{\sl Nucl. Phys.} {\bf B#2} [FS#1] (#3) #4}
\newcommand\CMP[3]{{\sl Commun. Math. Phys.} {\bf #1} (#2) #3}
\newcommand\PRD[3]{{\sl Phys. Rev.} {\bf D#1} (#2) #3}
\newcommand\PLA[3]{{\sl Phys. Lett.} {\bf #1A} (#2) #3}
\newcommand\PLB[3]{{\sl Phys. Lett.} {\bf #1B} (#2) #3}
\newcommand\JMP[3]{{\sl J. Math. Phys.} {\bf #1} (#2) #3}
\newcommand\PTP[3]{{\sl Prog. Theor. Phys.} {\bf #1} (#2) #3}
\newcommand\SPTP[3]{{\sl Suppl. Prog. Theor. Phys.} {\bf #1} (#2) #3}
\newcommand\AoP[3]{{\sl Ann. of Phys.} {\bf #1} (#2) #3}
\newcommand\RMP[3]{{\sl Rev. Mod. Phys.} {\bf #1} (#2) #3}
\newcommand\PR[3]{{\sl Phys. Reports} {\bf #1} (#2) #3}
\newcommand\FAP[3]{{\sl Funkt. Anal. Prilozheniya} {\bf #1} (#2) #3}
\newcommand\FAaIA[3]{{\sl Functional Analysis and Its Application} {\bf #1}
(#2) #3}
\def\TAMS#1#2#3{{\sl Trans. Am. Math. Soc.} {\bf #1} (#2) #3}
\def\InvM#1#2#3{{\sl Invent. Math.} {\bf #1} (#2) #3}
\def\AdM#1#2#3{{\sl Advances in Math.} {\bf #1} (#2) #3}
\def\PNAS#1#2#3{{\sl Proc. Natl. Acad. Sci. USA} {\bf #1} (#2) #3}
\newcommand\LMP[3]{{\sl Letters in Math. Phys.} {\bf #1} (#2) #3}
\newcommand\IJMPA[3]{{\sl Int. J. Mod. Phys.} {\bf A#1} (#2) #3}
\newcommand\TMP[3]{{\sl Theor. Mat. Phys.} {\bf #1} (#2) #3}
\newcommand\JPA[3]{{\sl J. Physics} {\bf A#1} (#2) #3}
\newcommand\JSM[3]{{\sl J. Soviet Math.} {\bf #1} (#2) #3}
\newcommand\MPLA[3]{{\sl Mod. Phys. Lett.} {\bf A#1} (#2) #3}
\newcommand\JETP[3]{{\sl Sov. Phys. JETP} {\bf #1} (#2) #3}
\newcommand\JETPL[3]{{\sl  Sov. Phys. JETP Lett.} {\bf #1} (#2) #3}
\newcommand\PHSA[3]{{\sl Physica} {\bf A#1} (#2) #3}
\newcommand\PHSD[3]{{\sl Physica} {\bf D#1} (#2) #3}
\newcommand\JPSJ[3]{{\sl J. Phys. Soc. Jpn.} {\bf #1} (#2) #3}

\newcommand\hepth[1]{{\sl hep-th/#1}}

\vspace*{-1cm}
\noindent
December, 1995 \hfill{INRNE-TH/95-15}\\
${}$ \hfill{UICHEP-TH/95-14} \\
${}$ \hfill{solv-int/9512008}
\begin{center}
{\Large {\bf Constrained KP Hierarchies: Darboux-B\"acklund Solutions\\
and Additional Symmetries \foot{To be published in
{\em ``Modern Trends in Quantum Field Theory''}, {\sl eds. A. Ganchev,
R. Kerner and I. Todorov}, Heron Press ({\sl Proceedings} of the second
Summer Workshop, Razlog/Bulgaria, Aug-Sept 1995).} }}
\end{center}
\vskip .3in
\begin{center}
{ H. Aratyn\footnotemark
\footnotetext{Work supported in part by U.S. Department of Energy,
contract DE-FG02-84ER40173}}

\par \vskip .1in \noindent
Department of Physics \\
University of Illinois at Chicago\\
845 W. Taylor St.\\
Chicago, IL 60607-7059, {\em e-mail}:
aratyn@uic.edu \\
\par \vskip .3in
E. Nissimov${}^{a),\, 3}$
and S. Pacheva${}^{a),\, b),}$\foot{Supported in part
by Bulgarian NSF grant {\em Ph-401}.}
\par \vskip .1in \noindent
${}^{a)}$ Institute of Nuclear Research and Nuclear Energy \\
Boul. Tsarigradsko Chausee 72, BG-1784 ~Sofia, Bulgaria \\
{\em e-mail}: emil@bgearn.bitnet, svetlana@bgearn.bitnet \\
and \\
${}^{b)}$ Department of Physics, Ben-Gurion University of the Negev \\
Box 653, IL-84105 $\;$Beer Sheva, Israel \\
{\em e-mail}: emil@bgumail.bgu.ac.il, svetlana@bgumail.bgu.ac.il
\end{center}
\vskip .3in

\begin{abstract}
We illustrate the basic notions of {\em additional non-isospectral
symmetries} and their interplay with the discrete {\em \DB transformations}
of integrable systems at the instance of {\em constrained
Kadomtsev-Petviashvili} (\cKP) integrable hierarchies. As a main application
we present the solution of discrete multi-matrix string models in terms of
Wronskian $\t$-functions of graded $SL(m,1)$ \cKP hierarchies.
\end{abstract}

\noindent
{\large {\bf 1. Introduction. ${\bf KP}$ as an Arch-Type Integrable System}}
\mskp
Integrable systems (for the basics, see refs. \ct{Faddeev,QISM,Olsha})
constitute an outstanding branch of theoretical physics
since they describe a vast variety of fundamental non-perturbative phenomena
ranging from $D=2$ (space-time dimensional)
nonlinear soliton physics and planar statistical mechanics to
string and membrane theories in high-energy elementary particle physics.
It turns out that, under plausible assumptions, a variety of physically
interesting theories in higher space-time dimensions can be reformulated as
lower-dimensional ($D=2$) integrable models which in the same time
possess {\em infinite-dimensional symmetries} and thus, as a rule, being
integrable (see, especially, the recent developments \ct{Seiberg-Witten}
related with integrability of Seiberg-Witten effective low-energy theory of
(extended) supersymmetric gauge theories).

Among the various infinite-dimensional symmetry groups and algebras playing
r\^{o}le in integrable field theories, a particularly distinguished place
belongs to the Lie algebra $W_{1+\infty}$ \ct{W-inf}
(specific ``large $N$ limit'' of Zamolodchikov's $W_N$ conformal algebras
\ct{Zam}, isomorphic to the Lie algebra of all
purely differential operators on the circle). It contains (together with its
supersymmetric extension) all previously known infinite-dimensional symmetry
algebras -- Virasoro and Kac-Moody. Also, it is precisely a Lie-algebraic
deformation of the infinite-dimensional generalization of the Virasoro
algebra -- the algebra of area-preserving diffeomorphisms.

Recently $\, W_{1+\infty}$ symmetries attracted broad interest as they
appeared naturally as inherent structures of models in different areas
of theoretical physics : theory of black holes and space-time
singularities, two-dimensional quantum gravity, nonlinear evolution
equations in higher dimensions, self-dual gravity,
$N=2$ superstring theory (refs. (a)--(d) in \ct{W-appl}),
quantum Hall effect \ct{Cappelli}.
All listed models possess, in one form or in
another, {\em exactly soluble} features, which naturally suggest an intimate
connection of {\em integrability} with $W_{1+\infty}$ algebra.
This claim may, furthermore, be substantiated by the realization that
$W_{1+\infty}$ algebra is a subalgebra of the algebra $\PsDA$ of arbitrary
pseudo-differential operators on the circle.
Already since the pioneering papers of Adler-Kostant-Symes and of the Faddeev's
school \ct{AKS} it was realized that $\PsDA$ forms the foundation
of completely integrable systems. In fact, it turns out that most known
integrable models (i.e. those admitting
Lax or ``zero curvature'' representation) can be associated with
specific coadjoint orbits of various subalgebras of $\PsDA$ or with
different Hamiltonian reductions thereof \ct{Faddeev}.

The generic integrable system based on $\PsDA$ symmetry algebra is
the Kadomtsev-Petviashvili (${\bf KP}$) integrable hierarchy \ct{Zakh,Dickey}
of soliton nonlinear evolution equations. Its name derives from the
fact that ${\bf KP}$ hierarchy contains the $D=2+1$ dimensional
nonlinear soliton ${\bf KP}$ equation which appeared originally in plasma
physics. In the last few years the main interest towards ${\bf KP}$ hierarchy
originates from
its deep connection with the statistical-mechanical models of random matrices
((multi-)matrix models) providing non-perturbative discretized formulation
of string theory \ct{integr-matrix}.

The purpose of the present talk is to provide a brief discussion of
the basic notions of {\em additional non-isospectral
symmetries} \ct{add-symm,Dickey} and their interplay with the
discrete {\em \DB transformations} \ct{DB}
of integrable systems within the ${\bf KP}$ integrable hierarchy
(being and arch-type integrable system, as pointed out above) and
its various constrained versions (\cKP hierarchies) relevant in discrete
multi-matrix string models. Furthermore, we show how to obtain the solution of
the latter in terms of Wronskian $\t$-functions of the \cKP hierarchies.

\mskp
{\large {\bf 2. ${\bf KP}$ Hierarchy: Pseudo-Differential Operator Formalism
and Hamiltonian \\
$\phantom{aaa}\!$Structures (${\bf W}$-Algebras)}}
\mskp
We describe the (generalized) ${\bf KP}$ integrable hierarchy in the
language of pseudo-differential operators (for a
background, see \ct{Dickey}). The main object is the pseudo-differential Lax
operator $L$ subject to an infinite set of evolution equations:
\be
L = D^m + \sum_{j=0}^{m-2} v_j D^j + \sum_{i \geq 1} u_i D^{-i}  \qquad ,\quad
\partder{}{t_l} L = \Sbr{L^{l\over m}_{(+)}}{L}
\lab{gen-KP}
\ee
Here, the coefficients of $L$ are (smooth) functions of $x \equiv t_1$ and
the higher time-evolution parameters $t_2 ,t_3 ,\ldots$ ; $D \equiv \pa_x$,
whereas the subscripts $(\pm)$ denote purely differential (purely
pseudo-differential) part of the corresponding pseudo-differential operators.
The flows $\partder{}{t_l}$ in \rf{gen-KP} commute (as vector fields on the
space of Lax operators \rf{gen-KP}) among
themselves which expresses the integrability of the ${\bf KP}$ system.

Within the Sato-Wilson dressing operator
formalism, with the following dressing expression for the generalized
${\bf KP}$ Lax operator \rf{gen-KP}:
\be
L= W D^m W^{-1} \qquad ,\quad
W \equiv 1 + \sum_{i \geq 1} w_i D^{-i}
\lab{dress-1}
\ee
the evolution equations for $L$ are equivalent to:
\be
\partder{}{t_l} W = - \( W D^l W^{-1} \)_{-} W
\lab{dress-2}
\ee

In what follows we shall also need the important notions of
{\em (adjoint) eigenfunctions} and {\em (adjoint) Baker-Akhiezer} functions.
The function $\Phi$ ($ \Psi$) is called \underbar{(adjoint) eigenfunction} for
the Lax  operator $L$ satisfying Sato's flow equations \rf{gen-KP} if its
flows are given by expression:
\be
\partder{\Phi}{t_l} = L^{l\over m}_{(+)} \Phi \qquad; \qquad
\partder{\Psi}{t_l} = - \( L^{*} \)^{l\over m}_{(+)} \Psi
\lab{eigenlax}
\ee
for the infinite many times $t_l$ \foot{Here and below, the superscript ``*''
on operators indicates pseudo-differential operator conjugation.}. If, in
addition, an (adjoint) eigenfunction satisfies the spectral equation:
\be
L\psi (\l ) = \l \psi (\l ) \;\; ,\;\;
\partder{}{t_l}\psi (\l ) =  L^{l\over m}_{(+)} \psi (\l ) \quad ; \quad
L^{*}\psi^{*} (\l ) = \l\psi^{*}(\l ) \;\; ,\;\;
\partder{}{t_l} \psi^{*} (\l ) = - \( L^{*} \)^{l\over m}_{(+)} \psi^{*}(\l )
\lab{BA-linsys}
\ee
$\psi^{(\ast )} (\l )$ is called \underbar{(adjoint) Baker-Akhiezer} (BA)
function.

The BA function of the generic Lax operator $L$ \rf{gen-KP} is
obtained from the BA function of the ``free'' Lax operator $L^{(0)} = D^m$ :
\be
\psi^{(0)} (\l ) = \exp \Bigl\{ \sum_{l \geq 1} t_l \l^{l \over m} \Bigr\}
\equiv e^{\xi (\{ t\}, \l )}
\lab{free-BA}
\ee
by applying the dressing operator $W$ \rf{dress-1} to \rf{free-BA} :
\be
\psi (\l ) = W e^{\xi (\{ t\}, \l )} =
\frac{\t (\{ t_l - 1/l\l^{l\over m}\} )}{\t (\{ t_l \} )} e^{\xi (\{ t\},\l )}
\lab{psi-tau}
\ee
The function $\t (\{ t_l \} )$ of all evolution parameters is called
\underbar{$\t$-function} of the (generalized) ${\bf KP}$ hierarchy and
by itself constitutes an alternative natural way to describe the pertinent
integrable system.

The ${\bf KP}$ system \rf{gen-KP} is endowed with bi-Hamiltonian
Poisson bracket structures (another expression of its integrability) which
results from the two compatible Hamiltonian
structures on the algebra of pseudo-differential operators $\PsDA$ \ct{STS83}.
The latter are given by:
\br
{\pbbr{\me{L}{X}}{\me{L}{Y}}}_1 \eq
- \llangle L \bv \left\lb X,\, Y \right\rb \rrangle
\lab{first-KP}\\
{\pbbr{\me{L}{X}}{\me{L}{Y}}}_2 \eq {\Tr}_A \( \( LX\)_{(+)} LY -
\( XL\)_{(+)} YL \)      \nonu  \\
&+& \!\! {1\over m}\int dx \, {\rm Res}\Bigl( \sbr{L}{X}\Bigr) \pa^{-1}
{\rm Res}\Bigl( \sbr{L}{Y}\Bigr) \qquad
\lab{second-KP}
\er
where the following notations are used.
$<\cdot \v \cdot >$ denotes the standard bilinear pairing in $\PsDA$ via
the Adler trace $\me{L}{X} = {\Tr}_A \( LX\)$ with
${\Tr}_A X = \int {\rm Res} X $. Here
$X,Y$ are arbitrary elements of the algebra of pseudo-differential
operators of the form $X = \sum_{k \geq - \infty} D^k X_k $ and similarly for
$Y$. The second term on the r.h.s. of \rf{second-KP} is a Dirac bracket term
originating from the second-class Hamiltonian constraint $v_{m-1}=0$ on $L$
\rf{gen-KP}.

In terms of the Lax coefficient functions
$v_{m-2},\ldots ,v_0 ,u_1 ,u_2 ,\ldots ,$ the first Poisson bracket structure
\rf{first-KP} takes the form of an infinite-dimensional Lie algebra which is
a direct sum of two subalgebras spanned by $\{ v_j\}$ and
$\{ u_i\}$, respectively. The latter is called $\Win1$-algebra \ct{W-inf}.
Its Cartan subalgebra contains the infinite set of (Poisson-)commuting
${\bf KP}$ integrals of motion ~$H_{l-1} = {1\over l} {\Tr}_A L^{l\over m}$
whose densities are expressed in terms of the $\t$-function \rf{psi-tau} as:
\be
\pa_x \partder{}{t_l} \ln \t = {\rm Res} L^{l\over m}
\lab{tau-L}
\ee

In turn, the second Poisson bracket structure \rf{second-KP} spans a nonlinear
(quadratic) algebra called $\hWinf (m)$ \ct{W-h-inf}, which is an
infinite-dimensional generalization of Zamolodchikov's $W_N$ conformal
algebras \ct{Zam}.

\mskp
{\large {\bf 3. Constrained ${\bf KP}$ Hierarchies. Free-Field Realizations}}
\mskp
Let us now turn our attention to a specific class of Hamiltonian reductions
of the full (generalized) ${\bf KP}$ system \rf{gen-KP} (\cKP hierarchies,
for short),
where the purely pseudo-differential part of the ${\bf KP}$ Lax operator is
parametrized through a {\em finite} number of functions (fields). To this end
let us recall the notion of (adjoint) eigenfunction of a Lax operator
\rf{eigenlax}.
As in \ct{chengs} let us consider the flow of a vector field $\pa_\a$ given by:
\be
\pa_{\a}L = \Sbr{L}{\sum_{i=1}^M \Phi_i D^{-1} \Psi_i}
\lab{ghostflo}
\ee
where $\Phi_i , \Psi_i$ are a set of $M$ independent (adjoint)
eigenfunctions of $L$ \rf{gen-KP}. Using the simple identity valid for any
differential operator $B_{(+)}$:
\be
{\Sbr{B_{(+)}}{\Phi_i D^{-1} \Psi_i}}_{-} = \( B_{(+)}\Phi_i \) D^{-1} \Psi_i
- \Phi_i D^{-1} \( B^{*}_{(+)}\Psi_i \)
\lab{Phi-Psi-id}
\ee
one can easily show that:
\be
\llb \pa_\a ,\partder{}{t_l} \rrb L = 0 \qquad ,\quad l=1,2,\ldots
\lab{comm}
\ee
Now, the constrained KP hierarchy (denoted as ${\sf cKP}_{m,M}$) is obtained by
identifying the ``ghost'' flow $\pa_{\a}$  with the isospectral flow
$\partder{}{t_m}$ which, upon comparison of \rf{ghostflo} with \rf{gen-KP},
implies the following constrained form of $L$:
\be
L \equiv L_{m,M} = L_{(+)} +  \sum_{i=1}^M \Phi_i D^{-1} \Psi_i
=D^m+ \sum_{j=0}^{m-2} v_j D^j + \sum_{i=1}^M \Phi_i D^{-1} \Psi_i
\lab{cKP-mM}
\ee
subject to the same Lax evolution equations as in \rf{gen-KP}. Moreover,
using again identity \rf{Phi-Psi-id} one finds that the functions
$\Phi_i , \Psi_i$ remain (adjoint) eigenfunctions of the
constrained Lax operator $L_{m,M}$ \rf{cKP-mM}.

As shown in ref.\ct{avoda}, the ${\sf cKP}_{m,M}$ hierarchies given by
\rf{cKP-mM} are equivalent to the so called ``multi-boson'' \cKP hierarchies
\ct{multi-b}:
\br
L_{m,M} = L_{(+)} +  \sum_{i=1}^M A^{(M)}_i \( D - B^{(M)}_i \)^{-1}
\( D - B^{(M)}_{i+1}\)^{-1} \cdots \( D - B^{(M)}_M \)^{-1}
\lab{iss-8a}  \\
A^{(M)}_k = (-1)^{M-k} \sum_{s=1}^k \Phi_s \frac{W\llb \Psi_M ,\ldots ,
\Psi_{k+1},\Psi_s \rrb}{W\llb \Psi_M ,\ldots ,\Psi_{k+1}\rrb}
\lab{iss-8b}\\
B^{(M)}_k = - \pa_x \ln \frac{W\llb \Psi_M ,\ldots ,
\Psi_{k+1},\Psi_k \rrb}{W\llb \Psi_M ,\ldots ,\Psi_{k+1}\rrb}
\lab{iss-8c}
\er
where
\be
W\llb f_1 ,\ldots ,f_k\rrb \equiv
\det {\Bigl\Vert} \pa_x^{i-1} f_j {\Bigr\Vert}
\lab{Wronskian}
\ee
denotes the standard Wronskian.

There is still another useful representation of the ${\sf cKP}_{m,M}$ Lax
operator as a ratio of two purely differential Lax operators
\ct{no2rabn1,office,rio}:
\br
L_{m,M} = L_{m+M} \( L_M\)^{-1}  \qquad ; \quad m , M \geq 1
\lab{lax-mM} \phanta
\lab{ratio-L}  \\
L_{m+M} \equiv \( D - b_{m+M}\) \( D - b_{m+M-1} \) \cdots \( D - b_1 \)
\quad , \quad
L_M \equiv
\( D - {\ti b}_{M}\) \( D - {\ti b}_{M-1} \) \cdots \( D - {\ti b}_1 \)
\lab{free-rep}
\er
where the coefficients $b_j , {\ti b}_j $ are subject to the constraint:
\be
\sum_{j=1}^{m+M} b_j - \sum_{l=1}^{M} {\ti b}_l = 0
\lab{b-constr}
\ee

As already proved in detail in ref.\ct{office}, the ${\sf cKP}_{m,M}$ Lax
operator $L=L_{m,M}$ obeys the same two compatible Poisson bracket structures
\rf{first-KP} and \rf{second-KP}, {\sl i.e.}, ${\sf cKP}_{m,M}$ hierarchies
are legitimate Hamiltonian reductions of the full (generalized) ${\bf KP}$
hierarchy. Moreover, the second Poisson bracket structure \rf{second-KP} in
terms of the coefficients $ $ \rf{free-rep} takes the form of free-field
Poisson bracket algebra:
\br
\pbbr{b_i (x)}{b_j (y)} \eq \( \d_{ij} - {1\over m}\) \d^{\pr} (x-y) \; ,
\qquad i,j =1,\ldots , m+M  \nonu \\
\pbbr{{\ti b}_k (x)}{{\ti b}_l(y)} \eq
- \( \d_{kl} + {1\over m}\) \d^{\pr} (x-y) \; ,\qquad k,l =1,\ldots ,M \nonu \\
\pbbr{b_i (x)}{{\ti b}_l (y)} \eq {1\over m} \d^{\pr} (x-y)
\lab{free-pb}
\er
which, as demonstrated in refs.\ct{Yu,office}, is precisely the Cartan
subalgebra of the graded $SL(m+M,M)$ Kac-Moody algebra. This latter property
justifies the alternative name of the constrained ${\sf cKP}_{m,M}$
hierarchies -- $SL(m+M,M)$ KP-KdV hierarchies.

In other words, \rf{lax-mM}--\rf{free-pb} provide via eq.\rf{second-KP}
explicit free-field realizations of the nonlinear $\hWinf (m)$ algebra.
Similar free-field realizations exist also for $\Win1$ -- the first ${\bf KP}$
Poisson bracket structure (see refs.\ct{multikp,office}).
\mskp
{\large {\bf 4. Additional Symmetries and \DB Transformations}}
\mskp
Let $L$ be again a pseudo-differential Lax operator of the full generalized KP
hierarchy \rf{gen-KP} (recall $x \equiv t_1$) and let $M$ be a
pseudo-differential operator ``canonically conjugated'' to $L$ such that:
\be
\Sbr{L}{M} = \one \quad , \quad
\partder{}{t_l} M = \Sbr{L^{l\over m}_{(+)}}{M}
\lab{L-M}
\ee
Within the Sato-Wilson dressing operator formalism \rf{dress-1}--\rf{dress-2}
the $M$-operator can be expressed in terms of dressing of the
``bare'' $M^{(0)}$ operator:
\br
M^{(0)} = \sum_{l \geq 1} \frac{l}{m} t_l D^{l-m} =
X_{(m)} + \sum_{l \geq 1} \frac{l+m}{m} t_{m+l} D^l
\lab{M-0} \\
X_{(m)} \equiv \sum_{l=1}^{m} \frac{l}{m} t_l D^{l-m}
\lab{X-m}
\er
conjugated to the ``bare'' Lax operator $L^{(0)} = D^m$, {\sl i.e.}:
\br
M \eq W M^{(0)} W^{-1} =
W X_{(m)} W^{-1} + \sum_{l \geq 1} \frac{l+m}{m} t_{m+l} L^{l\over m} =
\sum_{l \geq 0} \frac{l+m}{m} t_{m+l} L^{l\over m}_{(+)} + M_{-}
\lab{M-dress}  \\
M_{-} \eq W X_{(m)} W^{-1} - t_m -
\sum_{l \geq 1} \frac{l+m}{m} t_{m+l} \partder{}{t_l} W \, .\, W^{-1}
\lab{M--}
\er
where in \rf{M--} we used eqs.\rf{dress-2}.
Note that $X_{(m)}$ is a pseudo-differential operator satisfying
$\Sbr{D^m}{X_{(m)}} = \one$ .

On BA functions \rf{BA-linsys} the action of $M$ is as follows:
\be
M \psi (\l ) = \( \partder{}{\l} + \a_m (\l )\) \psi (\l )
\lab{M-psi}
\ee
where $\a_m (\l )$ is a function of $\l$ only\foot{The appearance of
$\a_m (\l )$ can be traced back to the ambiguity in the definition of the
dressing operator \rf{dress-1}: $ W \longrightarrow W W_0\,$ where
$W_0 = 1 + \sum_{i \geq 1} c_i D^{-i}\,$ with {\em constant} coefficients
$c_i$.}.

Since any eigenfunction $\Phi$ \rf{eigenlax} can be represented as a linear
``superposition'' of BA functions \rf{BA-linsys} :
\be
\Phi (\{ t\}) =
\int_{\Gamma} d\l \, \p (\l ) \psi (\l ,\{ t\})
\lab{Phi-psi}
\ee
(with an appropriate contour $\Gamma$ in the complex $\l$-plane, such that
the integral in \rf{Phi-psi} exists), eq.\rf{M-psi} implies that:
\be
M \Phi (\{ t\}) =
\int_{\Gamma} d\l \,\( -\partder{}{\l} + \a_m (\l )\)\p (\l )\psi (\l ,\{ t\})
\lab{M-Phi}
\ee

The so called {\em additional (non-isospectral) symmetries}
\ct{add-symm,Dickey}
are defined as vector fields on the space of ${\bf KP}$ Lax operators
\rf{gen-KP} or, alternatively, on the dressing operator \rf{dress-1},
through their flows as follows:
\be
{\bar \pa}_{k,n} L = - \Sbr{\( L^k M^n\)_{-}}{L} =
\Sbr{\( L^k M^n\)_{(+)}}{L} + n L^k M^{n-1}   \qquad , \qquad
{\bar \pa}_{k,n} W = - \( L^k M^n\)_{-} W
\lab{add-symm-L}
\ee
which {\em commute} with the usual ${\bf KP}$ flows $\partder{}{t_l}$
\rf{gen-KP}.

\vskip .1in

Let us now turn our attention to the notion of \DB (DB) transformations of
(generalized) ${\bf KP}$ hierarchy \rf{gen-KP} and its reductions --
${\sf cKP}_{m,M}$ hierarchies \rf{cKP-mM}, defined as follows \ct{DB,avoda} :
\br
{\wti L} &=& T L T^{-1} \equiv {\wti L}_{(+)} + {\wti L}_{-} \qquad ,\qquad
T \equiv \chi D \chi^{-1}
\lab{DB-1} \\
{\wti L}_{(+)} &=& {L}_{(+)} + \chi \(\pa_x \(\chi^{-1} L_{(+)}\chi\)_{\geq 1}
D^{-1}\) \chi^{-1}
\lab{DB-2}
\er
\br
{\wti L}_{-} &=& {\wti \Phi}_0 D^{-1} {\wti \Psi}_0 +
\chi D \chi^{-1} L_{-} \chi D^{-1} \chi^{-1} \quad \( \; =
{\wti \Phi}_0 D^{-1} {\wti \Psi}_0 +
\sum_{i=1}^{M} {\wti \Phi}_i D^{-1} {\wti \Psi}_i
\quad {\rm for} \;\; L = L_{m,M} \; \)
\lab{DB-3} \\
{\wti \Phi}_0 &=& \( \chi D \chi^{-1} L \) \chi \equiv TL\chi \qquad ,\qquad
{\wti \Psi}_0 = \chi^{-1}
\lab{DB-4} \\
{\wti \Phi}_i &=& \chi \pa_x \( \chi^{-1} \Phi_i \) \qquad ,\qquad
{\wti \Psi}_i = - \chi^{-1} \pa_x^{-1} \( \Psi_i \chi \)
\lab{DB-5}
\er
where $\chi$ is an (non-BA) eigenfunction of $L$. The DB-transformed Lax
operator \rf{DB-1} satisfies the same flow equations w.r.t. $t_l$ as in
\rf{gen-KP}:
$\partder{}{t_l} {\wti L} = \Sbr{{\wti L}^{l\over m}_{(+)}}{{\wti L}}$~
due to the simple identity valid for any pseudo-differential operator $B$
\be
\( \chi D \chi^{-1} B \chi D^{-1} \chi^{-1} \)_{(+)} =
\chi D \chi^{-1}  B_{(+)}  \chi D^{-1} \chi^{-1}
-  \chi \pa_x \(\chi^{-1} (B_{(+)} \chi) \)  D^{-1} \chi^{-1}
\lab{B-chi}
\ee
(and using the fact that $\chi$ is an eigenfunction of $L$). Moreover,
eq.\rf{DB-3} shows that, in order to preserve the ${\sf cKP}_{m,M}$ form of
the DB-transformed Lax operator ${\wti L}={\wti L}_{m,M}$, we have to choose
$\chi = \Phi_{i_0}$ where $\Phi_{i_0}$ is any one of the eigenfunctions of the
initial $L=L_{m,M}$ \rf{cKP-mM}.

One can generalize \rf{DB-1}--\rf{DB-5} for successive \DB transformations
on the initial $L=L_{m,M}\equiv L^{(0)}$ as follows. Within each subset of
$m$ successive steps we can perform the DB transformations w.r.t. the $m$
different eigenfunctions of \rf{cKP-mM}.
Repeated use of the following important
composition formula for Wronskians \ct{Wronski} :
\be
T_k \, T_{k-1}\, \cdots\, T_1 (f ) \; =\; { W_{k} (f) \over W_k}
\lab{iw}
\ee
where
\br
T_j = { W_{j} \over W_{j-1} } D { W_{j-1} \over W_{j} } =
\( D + \( \ln { W_{j-1} \over W_{j} } \)^{\pr} \) \quad;\quad W_{0}=1
\lab{transf} \\
W_k \equiv W_k \lb \psi_1, \ldots ,\psi_k \rb  \quad ,\quad
W_{k-1} \(f \)\equiv W_{k} \lb \psi_1, \ldots ,\psi_{k-1}, f\rb
\lab{W-def}
\er
and employing short-hand notations:
\be
T^{(k)}_i \equiv \Phi^{(k)}_i D \(\Phi^{(k)}_i\)^{-1} \qquad ;\qquad
\chi^{(s)}_i \equiv \( L^{(0)}\)^s \Phi^{(0)}_i \quad ,\;\; i=1,\ldots ,m
\lab{defchi-i}
\ee
where the upper indices in brackets indicate the order of the corresponding
DB step, yields the following generalization of \rf{DB-4}--\rf{DB-5} (below
$1 \leq l \leq m$) :
\br
\Phi_i^{(km+l)} = T^{(km+l-1)}_l \ldots T^{(km)}_1
T^{(km-1)}_m \ldots T^{((k-1)m)}_1 \ldots T^{(m-1)}_m \ldots T^{(0)}_1
\chi^{(k_{\pm})}_i   \nonu  \\
= \frac{W\llb \Phi^{(0)}_1 ,\ldots ,\Phi^{(0)}_m ,
\chi^{(1)}_1,\ldots ,\chi^{(1)}_m ,\ldots ,\chi^{(k-1)}_1,\ldots
,\chi^{(k-1)}_m
 ,
\chi^{(k)}_1,\ldots ,\chi^{(k)}_l , \chi^{(k_{\pm})}_i \rrb}{W\llb
\Phi^{(0)}_1 ,\ldots ,\Phi^{(0)}_m ,
\chi^{(1)}_1,\ldots ,\chi^{(1)}_m ,\ldots ,\chi^{(k-1)}_1,\ldots
,\chi^{(k-1)}_m
 ,
\chi^{(k)}_1,\ldots ,\chi^{(k)}_l \rrb}
\lab{pchi-a-1}  \\
\chi^{(k_{+})}_i \equiv \chi^{(k+1)}_i \quad {\rm for} \;\; 1 \leq i \leq l
\qquad ; \qquad
\chi^{(k_{-})}_i \equiv \chi^{(k)}_i \quad {\rm for} \;\; l+1 \leq i \leq m
\nonu
\er
Correspondingly, for the $\t$ function \rf{psi-tau} after $km+l$ steps of
successive DB transformations we get:
\br
\frac{\t^{(km+l)}}{\t^{(0)}} = \Phi^{(km+l-1)}_l \ldots \Phi^{(km)}_1
\Phi^{(km-1)}_m \ldots \Phi^{((k-1)m)}_1 \ldots
\Phi^{(m-1)}_m \ldots \Phi^{(0)}_1
\nonu  \\
= W\llb \Phi^{(0)}_1 ,\ldots ,\Phi^{(0)}_m ,
\chi^{(1)}_1,\ldots ,\chi^{(1)}_m,\ldots ,\chi^{(k-1)}_1,\ldots ,\chi^{(k-1)}_m
 ,
\chi^{(k)}_1,\ldots ,\chi^{(k)}_l \rrb
\lab{tauok-1}
\er

\vskip .1in

We now formulate the main result of this section -- the condition for
{\em compatibility} between additional-symmetry flows \rf{add-symm-L} and \DB
transformations \rf{DB-1}.

Let $\Phi$ be an eigenfunction of $L$ defining a \DB transformation,
{\sl i.e.} :
\be
\partder{}{t_l} \Phi = L^{l\over m}_{(+)} \Phi  \quad ,\quad
{\wti L} = \(\Phi D \Phi^{-1}\)\, L\, \(\Phi D^{-1} \Phi^{-1}\)
\lab{gen-L-DB}
\ee
or, in terms of dressing operator:
\be
{\wti W} = \(\Phi D \Phi^{-1}\)\, W\, D^{-1}
\lab{gen-W-DB}
\ee
Then the DB-transformed $M$ operator (cf. \rf{M-dress}) acquires the form:
\br
{\wti M} = \(\Phi D \Phi^{-1}\)\, M\, \(\Phi D^{-1} \Phi^{-1}\) =
\sum_{l \geq 0} \frac{l+m}{m} t_{m+l} {\wti L}^{l\over m}_{(+)} + {\wti M}_{-}
\lab{gen-M-DB}  \\
{\wti M}_{-} = {\wti W} {\wti X}_{(m)} {\wti W}^{-1} - t_m -
\sum_{l \geq 1} \frac{l+m}{m} t_{m+l}\partder{}{t_l}{\wti W}\,.\,{\wti W}^{-1}
\lab{M--ti}
\er
where ${\wti X}_{(m)} = D X_{(m)} D^{-1}$ with $X_{(m)}$ as in \rf{X-m}.
Clearly ${\wti X}_{(m)}$, like $X_{(m)}$, is
also admissible as canonically conjugated to $D^m$.

The DB-transformed BA function reads in accordance with \rf{gen-W-DB}:
\be
{\wti \psi}(\l ) = \l^{-{1\over m}} \Phi \pa_x \( \Phi^{-1} \psi (\l )\)
\lab{DB-BA}
\ee
and the DB-transformed $M$-operator acts on it as:
\be
{\wti M} {\wti \psi} (\l ) = \( \partder{}{\l} + {\wti \a_m} (\l )\)
{\wti \psi} (\l )      \quad , \quad
{\wti \a_m} (\l ) = \a_m (\l ) + {1\over m} \l^{-1}
\lab{M-psi-DB}
\ee

Taking into account \rf{gen-L-DB}, we arrive at the following important
\mskp
\prop
{\em Additional symmetry flows \rf{add-symm-L} commute with
\DB transformations \rf{gen-L-DB}--\rf{gen-M-DB}, i.e.
\be
{\bar \pa}_{k,n} {\wti L} =
- \Sbr{\( {\wti L}^k {\wti M}^n\)_{-}}{{\wti L}}  \qquad ,\qquad
{\bar \pa}_{k,n} {\wti W} =
- \( {\wti L}^k {\wti M}^n\)_{-} {\wti W}
\lab{AS-DB-compat}
\ee
if and only if the {\rm DB-generating} eigenfunction $\Phi$
transforms under the additional symmetries as:
\be
{\bar \pa}_{k,n} \Phi = \( L^k M^n\)_{(+)} \Phi
\lab{add-symm-Phi-0}
\ee
}

Motivated by applications to (multi-)matrix models (see next sections and
ref.\ct{oakpark}), one can require
invariance under some of the additional-symmetry flows, {\sl e.g.},
under the lowest one ${\bar \pa}_{0,1}$ known as ``string-equation''
constraint in the context of (multi-)matrix models:
\br
{\bar \pa}_{0,1} L = 0  \qquad \to \quad  \Sbr{M_{(+)}}{L} = - \one
\qquad ; \qquad
{\bar \pa}_{0,1} \Phi = 0  \qquad \to \quad  M_{(+)} \Phi = 0
\lab{01-constr-L}
\er
Eqs.\rf{01-constr-L}, using second
eq.\rf{L-M},\rf{M-dress}--\rf{Phi-psi} and
\rf{add-symm-Phi-0}, lead to the following constraints for $L$, the BA
function $\psi (\l )$ and the
DB-generating eigenfunction $\Phi$ of $L$, respectively:
\br
\sum_{l \geq 1} \frac{l+m}{m} t_{m+l} \partder{}{t_l} L +
\Sbr{t_1}{L}\, \d_{m,1} = - \one
\lab{L-constr} \\
\( \sum_{l \geq 1} \frac{l+m}{m} t_{m+l} \partder{}{t_l} +
t_m - \a_m (\l )\) \psi (\l ) = \partder{}{\l} \psi (\l )
\lab{psi-constr} \\
\( \sum_{l \geq 1} \frac{l+m}{m} t_{m+l} \partder{}{t_l} + t_m \) \Phi =  0
\qquad  \to \qquad   \Phi (\{ t\}) =
\int_{\Gamma} d\l \, e^{\int \a (\l )} \psi (\l ;\{ t\})
\lab{P-constr}
\er

Now let us recall the formula \rf{tauok-1} for the $\t$-function ratio for
${\sf cKP}_{m,M}$ hierarchies
subject to successive DB transformations. Noticing that the
eigenfunctions $\Phi^{(k)}$ of the DB-transformed Lax operators $L^{(k)}$
satisfy the {\em same} constraint eq.\rf{P-constr} irrespective of the
DB-step $k$, we arrive at the following result (``string-equation''
constraint on the $\t$-functions) :
\mskp
\prop
{\em The Wronskian $\t$-functions \rf{tauok-1} of the ${\sf cKP}_{m,M}$
hierarchies satisfy the constraint equation:
\be
\( \sum_{l \geq 1} \frac{l+m}{m} t_{m+l} \partder{}{t_l} + nt_m \)
\frac{\t^{(n)}}{\t^{(0)}} = 0
\lab{tau-M-constr}
\ee
}

\underbar{Example:} It is well-known that the discrete one-matrix model
can be associated to the following chain of the Lax operators
connected via DB transformations:
\br
L^{(k+1)} \eq \(\Phi^{(k)}  D {\Phi^{(k)} }^{-1}\)  \; L^{(k)}
 \; \(\Phi^{(k)}  D^{-1} {\Phi^{(k)} }^{-1}\)
= D + \Phi^{(k+1)} D^{-1} \Psi^{(k+1)}
\lab{lkplus} \\
\Phi^{(k+1)} \eq \Phi^{(k)} \( \ln \Phi^{(k)}\)^{\pr \pr} + \(\Phi^{(k)}\)^2
\Psi^{(k)} \quad ,\quad \Psi^{(k+1)} = \(\Phi^{(k)}\)^{-1} \lab{pkplus}
\er
where
\be
\Phi^{(n)} = { W_{n+1} \lb \p , \pa \p, \ldots , \pa^n \p \rb \over
W_n \lb \p , \pa \p, \ldots , \pa^{n-1} \p \rb}  \qquad , \qquad
\frac{\t^{(n)}}{\t^{(0)}} = W_n \lb \p , \pa \p, \ldots , \pa^{n-1} \p \rb
\lab{phik}
\ee
with $\p =  \int d \l   \exp \(\sumi{k=1} t_k \l^k  \)$.
The above proposition (with $m=1$) coincides perfectly with the
``string-equation''
${\cal L}_{-1}^{(N)} W_N \lb \p, \pa \p, \ldots , \pa^{N-1} \p \rb =0$, with
${\cal L}_{-1}^{(N)} = \sumi{k=2} k t_k \partder{}{t_{k-1}} + N t_1$.

\mskp
{\large {\bf 5. Multi-Matrix Models as ${\bf cKP_{m,1}}$ Hierarchies}}
\mskp
The partition function of the multi-matrix ($q$-matrix) string model reads:
\be
Z_N \lb \{ t^{(1)}\},\ldots ,\{ t^{(q)}\},\{ g\} \rb =
\int dM_1 \ldots dM_q \exp -\lcurl
\sum_{\a =1}^q \sum_{r_\a =1}^{p_\a} t^{(\a )}_{r_\a} \Tr M_\a^{r_\a} +
\sum_{\a =1}^{q-1} g_{\a ,\a +1} \Tr M_\a M_{\a +1} \rcurl
\lab{ZN-qM}
\ee
where $M_\a$ are Hermitian $N \times N$ matrices, and the orders of the
matrix ``potentials'' $p_{\a}$ may be finite or infinite.
In refs.\ct{BX} it was shown\foot{See also refs.\ct{BX-2} and the
lecture of Prof. L.Bonora in the present volume.}
that, via the method of generalized orthogonal
polynomials \ct{ortho-poly}, one associates to \rf{ZN-qM} generalized
Toda-like lattice systems subject to specific constraints, so that
$Z_N$ and its derivatives w.r.t. the coupling parameters can be expressed in
terms of solutions of the underlying Toda-like discrete integrable hierarchy
where $\{ t^{(1)}\},\ldots ,\{ t^{(q)}\}$ play the role of ``evolution''
parameters. This Toda-like discrete integrable hierarchy differs from the
full generalized Toda lattice hierarchy \ct{U-T} in that the
associated Toda matrices in the first hierarchy are semiinfinite and
contain in general {\em finite} number of non-zero diagonals.

It turns out that, in order to identify the continuum \cKP integrable hierarchy
which provides the exact solution for \rf{ZN-qM}, we need the following subset
of the associated linear system and the corresponding Lax
(``zero-curvature'') representation from the Toda-like lattice system \ct{BX} :
\br
{Q(1)}_{nm} \psi_m = \l \psi_n  \quad , \quad
\partder{}{t^{(1)}_r} \psi_n = - \( {Q(1)}^r_{-}\)_{nm} \psi_m  \quad , \quad
\partder{}{t^{(q)}_s} \psi_n = - \( {Q(q)}^s_{-}\)_{nm} \psi_m
\lab{L-q-1-2} \\
\partder{}{t^{(1)}_r} Q(1) = \llb {Q(1)}^r_{(+)} , Q(1) \rrb  \quad , \quad
\partder{}{t^{(q)}_s} Q(1) = \llb Q(1) , {Q(q)}^s_{-} \rrb
\lab{L-q-3} \\
\partder{}{t^{(1)}_r} Q(q) = \llb Q^r_{(+)} , Q(q) \rrb  \quad , \quad
\partder{}{t^{(q)}_s} Q(q) = \llb Q(q) , {Q(q)}^s_{-} \rrb
\lab{L-q-4}
\er
In what follows it is convenient to introduce the short-hand notations:
\be
t_r \equiv t^{(1)}_r \quad ,\;\; r=1,\ldots ,p_1 \quad ; \quad
\tit_s \equiv t^{(q)}_s \quad ,\;\; s=1,\ldots ,p_q  \qquad ; \qquad
Q \equiv Q(1) \quad ,\quad \tQ \equiv Q(q)
\lab{t-q}
\ee
Further, there is a series of additional constraints
(``coupling conditions'') relating $Q \equiv Q(1)$ and $\tQ \equiv Q(q)$.
In the two-matrix model case ($q=2$) their explicit form is:
\br
-g \llb Q , \tQ \rrb = \one
\lab{string-eq-a} \\
Q_{(-)} = - \sum_{s=1}^{p_2 -1} \frac{(s+1)}{g} {\ti t}_{s+1} \tQ^s_{(-)}
 - {1\over g} {\ti t}_1 \one
\lab{string-eq-b} \\
\tQ_{(+)} = - \sum_{r=1}^{p_1 -1} \frac{(r+1)}{g} {t}_{r+1} Q^r_{(+)}
 - {1\over g} {t}_1 \one
\lab{string-eq-c}
\er
Here the subscripts $-/+$ denote lower/upper
triangular parts, whereas $(+)/(-)$ denote upper/lower triangular plus
diagonal parts. In the higher ($q \geq 3$) multi-matrix case the
``coupling conditions'' have much more intricate form (involving also the
``intermediate'' $Q(2),\ldots ,Q(q-1)$ matrices).
However, their explicit form will not be needed to find the solution for
$Z_N$ \rf{ZN-qM} since we will be able to extract the
relevant information solely from the discrete Lax system \rf{L-q-3}--\rf{L-q-4}
and the relations expressing $Q \equiv Q(1),\,\tQ \equiv Q(q)$ in
terms of orthogonal polynomial factors (see eqs.\rf{abR-h} below).

The parametrization for the matrix elements of the Jacobi matrices
$Q \equiv Q(1)$ and $\tQ \equiv Q(q)$ is as follows:
\br
Q_{nn} \equiv {Q(1)}_{nn} = a_0 (n) \quad , \quad
{Q(1)}_{n,n+1} \equiv Q_{n,n+1} =1   \phanta   \nonu  \\
{Q(1)}_{n,n-k} \equiv Q_{n,n-k} = a_k (n) \quad k=1,\ldots , m(1) \quad, \quad
m(1) = (p_q -1) \ldots (p_2 -1)   \nonu   \\
{Q(1)}_{nm} \equiv Q_{nm} = 0 \quad {\rm for} \;\;\; m-n \geq 2 \;\; ,\;\;
n-m \geq m(1) +1    \phanta
\lab{param-q-1}  \\
{Q(q)}_{nn} \equiv \tQ_{nn} = b_0 (n) \quad , \quad
{Q(q)}_{n,n-1} \equiv \tQ_{n,n-1} = R_n     \phanta    \nonu  \\
{Q(q)}_{n,n+k} \equiv \tQ_{n,n+k} = b_k (n) R_{n+1}^{-1} \ldots R_{n+k}^{-1}
\quad k=1,\ldots , m(q)   \quad , \quad
m(q) = (p_{q-1} -1) \ldots (p_1 -1)   \nonu  \\
{Q(q)}_{nm} \equiv \tQ_{nm} = 0 \quad {\rm for} \;\;\; n-m \geq 2 \;\; ,\;\;
m-n \geq m(q) +1     \phanta
\lab{param-q-2}
\er

In terms of the $Q \equiv Q(1),\,\tQ \equiv Q(q)$ matrix elements the
partition function \rf{ZN-qM} is expressed in the following way \ct{BX}:
\br
Z_N = const ~\prod_{n=0}^{N-1} h_n
\lab{ZN-h} \\
a_0 (n) = \partder{}{t_1} \ln h_n \quad ,\quad
b_0 (n) = \partder{}{\tit_1} \ln h_n \quad ,\quad
R_n = \frac{h_n}{h_{n-1}}        \nonu  \\
\partder{}{t_r} \ln h_n = Q^r_{nn} \quad ,\quad
\partder{}{\tit_s} \ln h_n = \tQ^s_{nn}
\lab{abR-h} \\
\partder{}{t_1} \ln Z_N = \sum_{n=0}^{N-1} a_0 (n) \quad ,\quad
\partder{}{\tit_1} \ln Z_N = \sum_{n=0}^{N-1} b_0 (n)  \nonu  \\
\partder{}{t_r} \ln Z_N = \sum_{n=0}^{N-1} Q^r_{nn}   \quad ,\quad
\partder{}{\tit_s} \ln Z_N = \sum_{n=0}^{N-1} \tQ^s_{nn}
\lab{ZN-ab}
\er
where $h_n$ are the normalization
factors in the nonlocally generalized orthogonal polynomial formalism
\ct{ortho-poly} (using notations \rf{t-q}) :
\br
h_n \d_{nm} = \int_{\Gamma} \int_{\Gamma} d\l d\m \,
P_n (\l ) \exp \Bigl\{ \sum_{r=1}^{p_1} \l^r t_r \Bigr\}
\rho (\l ,\m ; \{ t^{\pr\pr}\}, \{ g\})
\Bigl\{ \sum_{s=1}^{p_q} \m^s \tit_s \Bigr\} {\wti P}_m (\m )
\lab{h-q-n} \\
\rho (\l ,\m ; \{ t^{\pr\pr}\}, \{ g\}) =
\int_{\Gamma} \prod_{\a =2}^{q-1} d\n_\a \, \exp \Bigl\{
\sum_{\a =2}^{q-1} \sum_{r_\a =1}^{p_\a} t^{(r_\a )}_{r_\a} \n_\a^{r_\a} +
\sum_{\a =2}^{q-2} g_{\a ,\a +1} \n_\a \n_{\a +1} +
g_{12} \l \n_2 + g_{q-1,q} \n_{q-1} \m \Bigr\}
\lab{rho-q} \\
\{ t^{\pr\pr}\} \equiv \( t^{(2)},\ldots ,t^{(q-1)}\)  \phanta \nonu
\er

As in the case of two-matrix model \ct{office,avoda}, using the lattice
equations of motion (eqs.\rf{L-q-3}--\rf{L-q-4} for $r=1,s=1$) we obtain the
following important:

\prop
{\em The matrix elements of $Q \equiv Q(1)$ are completely expressed in terms
of the matrix elements of $\tQ \equiv Q(q)$ through the relations:
\be
Q_{(-)} = \sum_{s=0}^{m(1)} \a_s \tQ^s_{(-)} \qquad ,\qquad
Q^{s\over {m(1)}}_{(-)} = \sum_{\s =0}^{s} \g_{s\s} \tQ^{\s}_{(-)}
 \quad , \;\; s=0,1,\ldots ,m(1)
\lab{tatko-q-s}
\ee
where the coefficients $\a_0 ,\g_{s,0}$ are $t_1$-independent,
whereas the coefficients $\a_s ,\g_{s\s}$ with $\s \geq 1$ are independent of
$t_1, {\ti t}_1$. All $\g_{s\s}$ are expressed through $\a_s\equiv \g_{m(1),s}$
solely:
\br
\g_{ss} = \( \g_{11}\)^s \;, \;
\g_{s,s-1} = s \(\g_{11}\)^{s-1} \g_{10}  \;, \;
\g_{s,s-2} = \(\g_{11}\)^{s-2} \llb \frac{s(s-1)}{2} \(\g_{10}\)^2 +
s \( \frac{\g_{31}}{3\g_{11}} - \g_{10}^2 \) \rrb
\lab{tatko-q-2-a}  \\
\g_{11} = \( \a_{m(1)}\)^{1\over {m(1)}} \quad ,\quad
\g_{10} = \frac{\a_{m(1)-1}}{m(1) \( \a_{m(1)}\)^{{m(1)-1}\over {m(1)}}}
\nonu  \\
\frac{\g_{31}}{3\g_{11}} - \g_{10}^2 = {1\over {m(1)}}
\llb \frac{\a_{m(1)-2}}{\( \a_{m(1)}\)^{{m(1)-2}\over {m(1)}}} -
\frac{m(1)-1}{2m(1)}\,
\frac{\a^2_{m(1)-1}}{\( \a_{m(1)}\)^{{2(m(1)-1)}\over {m(1)}}} \rrb
\lab{tatko-q-2}
\er
}
\mskp
In the two-matrix model the explicit form of the coefficients $\a_s$ reads:
$\a_s = - \frac{(s+1)}{g}\tit_{s+1}$.

Similarly, we have the dual statement
with the r\^{o}les of $Q \equiv Q(1)$ and $\tQ \equiv Q(q)$ interchanged.

As an important consequence of \rf{tatko-q-s}, let us take
its diagonal $00$-part and use the last eq.\rf{abR-h} which yields:
\be
\partder{}{t_1} h_0 =
\( \sum_{s=1}^{m(1)} \a_s \partderh{}{\tit_1}{s} + \a_0 \) h_0
\lab{h-0-constr}
\ee
This equation is the only remnant of the constraints (``coupling conditions'')
on the multi-matrix model $Q$-matrices which will be used in the sequel.

Based on our experience with the two-matrix model \ct{office,avoda}, it
turns out natural to introduce the fractional power of $Q \equiv Q(1)$ :
\be
\hQ = Q^{1\over {m(1)}} \equiv {Q(1)}^{1\over {m(1)}}
\lab{Q-hat-q}
\ee
whose parametrization closely resembles that of $\tQ \equiv Q(q)$
\rf{param-q-2} :
\br
\hQ_{nn} = \hb_0 (n) \quad , \quad \hQ_{n,n-1} = \hR_n \quad , \quad
\hQ_{n,n+k} = \hb_k (n) \hR_{n+1}^{-1} \ldots \hR_{n+k}^{-1}
\quad k \geq 1    \nonu  \\
\hQ_{nm} = 0 \quad {\rm for} \;\;\; n-m \geq 2
\lab{param-2-h}
\er
{}From eqs.\rf{tatko-q-s} we find the following relation between the matrix
elements of $\hQ \equiv Q(1)^{1\over {m(1)}}$ and $\tQ$ :
\be
{\hat R}_n = \g_{11} R_n \quad , \quad {\hat b}_0 (n) = \g_{11} b_0 (n) +
\g_{10} \quad ,\quad {\hat b}_1 (n) = \g_{11}^2 b_1 (n) +
\frac{\g_{31}}{3\g_{11}} - \g_{10}^2
\lab{tatko-1}
\ee
etc., with $\g$-coefficients as in \rf{tatko-q-2-a}--\rf{tatko-q-2}.

In order to identify the continuum \cKP hierarchy associated with the general
$q$-matrix model,
as a first step we reexpress, using \rf{tatko-q-s}, the Toda-like lattice
hierarchy \rf{L-q-1-2}--\rf{L-q-4} as a single set of flow equations for
$\hQ \equiv {Q(1)}^{1\over {m(1)}}$ :
\br
\hQ^{m(1)}_{nm} {\psi}_m = \l {\psi}_n  \quad , \quad
\partder{}{{\htt}_s} {\psi}_n = - \({\hQ}^s_{(-)}\)_{nm} {\psi}_m
\lab{L-q-hat} \\
\partder{}{{\htt}_s} \hQ = \llb \hQ , \hQ^{s}_{-} \rrb \quad
\;,\; \;\; s=1,\ldots,p_q , 2m(1), 3m(1), \ldots, p_1 m(1)
\lab{L-3-q-h} \\
t_r \equiv {\hat t}_{rm(1)} \quad {\rm for} \;\; r=1,\ldots , p_1
\lab{identif-q}
\er
Here, as in the two-matrix case \ct{office,avoda}, we have introduced a new
subset of evolution parameters $\lcurl {\hat t}_s \rcurl$ instead of
$\lcurl {\ti t}_s \equiv t^{(q)}_s \rcurl$ defined as:
\be
\partder{}{{\hat t}_s} = \sum_{\s =1}^s \g_{s\s} \partder{}{{\ti t}_{\s}}
\quad ,\;\; s=1,\ldots, m(q)      \lab{tatko-q-a}
\ee
with the same $\g_{s\s}$ as in \rf{tatko-q-s}.
As a second step, one employs the Bonora-Xiong procedure \ct{BX} to get from
the discrete Lax system \rf{L-q-hat}--\rf{L-3-q-h} an equivalent continuum Lax
system associated with a fixed lattice site $n$, where the continuum space
coordinate is $x \equiv \htt_1$. Namely, the latter
continuum integrable system is obtained by writing
eqs.\rf{L-q-hat} in more detail using the
parametrization \rf{param-q-1},\rf{param-2-h} :
\br
\l \psi_n &=& \psi_{n+1} + a_0 (n) \psi_n + \sum_{k=1}^{p_2 -1}a_k
(n)\psi_{n-k}
\lab{3-1} \\
\partder{}{\htt_1} \psi_n &=& - \hR_n \psi_{n-1}
\lab{3-3}
\er
and further using \rf{3-3} to express $\psi_{n \pm\ell}$ in terms of
$\psi_n$ at a fixed lattice site $n$ in eq.\rf{3-1} and the higher evolution
eqs.\rf{L-q-hat} (for $s \geq 2$).
Upon operator conjugation and an appropriate similarity transformation,
it acquires the form (as before $x \equiv \htt_1$) :
\br
\partder{}{{\hat t}_s} L(n) =
\Sbr{\( L^{s\over {m(1)}}(n)\)_{(+)}}{L (n)}
\quad , \;\; s=1,\ldots,p_q , 2m(1), 3m(1), \ldots ,p_1 m(1)
\lab{Lax-q-2M}  \\
L(n) = D_x^{m(1)} + m(1) {\hb}_1 (n) D_x^{m(1)-2} + \cdots +
{\hR}_{n+1} \( D_x - {\hb}_0 (n) \)^{-1}
\lab{Ln-q}
\er
where $\hb_{0,1}(n), \hR_{n+1}$ are the matrix elements of $\hQ$
\rf{param-2-h},\rf{tatko-1}. Rewriting \rf{Ln-q} in the equivalent
``eigenfunction'' form:
\br
L(n) = D_x^{m(1)} + m(1) {\hb}_1 (n) D_x^{m(1)-2} + \cdots +
\Phi (n+1) D_x^{-1} \Psi (n+1)
\lab{Ln1} \\
\Phi (n+1) \equiv \hR_{n+1} \exp \Bigl\{ \int \hb_0 (n) \Bigr\} \quad ,\quad
\Psi (n+1) \equiv \exp \Bigl\{ -\int \hb_0 (n) \Bigr\}
\lab{Phi-n}
\er
and comparing with \rf{cKP-mM}, we identify the continuum integrable hierarchy
\rf{Lax-q-2M}, describing equivalently the discrete multi-matrix model,
as a constrained $\cKP_{m(1),1}$ hierarchy.

\mskp
{\large {\bf 6. Partition Functions of Multi-Matrix Models: \DB Solutions}}
\foot{A more detailed presentation of the material in this section will
appear in ref.\ct{oakpark}.}
\mskp
Exactly as in the two-matrix case \ct{office,avoda}, lattice shifts
$n \to n+1$ in the underlying discrete Toda lattice system, described by
\rf{L-q-1-2}--\rf{Phi-n}, generate \DB transformations in the continuum
$\cKP_{m(1),1}$ hierarchy \rf{Lax-q-2M}--\rf{Ln-q}. This is due to the fact
that the latter continuum hierarchy preserves its form for any value of the
discrete label $n$. The solutions
for the eigenfunctions and $\t$-functions at each successive step of \DB
transformation is given explicitly, as particular cases of
eqs.\rf{pchi-a-1}--\rf{tauok-1}, by:
\br
\Phi (n) = \frac{W_{n+1} \llb \Phi (0), L(-1)\Phi (0), \ldots, L(-1)^n \Phi (0)
\rrb}{W_n \llb \Phi (0), L(-1)\Phi (0), \ldots, L(-1)^{n-1} \Phi (0) \rrb}
\lab{Phi-W-n} \\
\frac{\t (n)}{\t (-1)} = \prod_{j=0}^n \Phi (j) =
W_{n+1} \llb \Phi (0), L(-1)\Phi (0), \ldots, L(-1)^n \Phi (0) \rrb
\lab{tau-W-n}   \\
\partder{}{\htt_s} \Phi (0) = \( L(-1)\)^{s\over{p_2 -1}}_{(+)} \Phi (0)
\quad , \;\;\; s=1,\ldots,p_2 , 2(p_2 -1), 3(p_2 -1), \ldots, p_1 (p_2 -1)
\lab{P0-eigen}
\er
where everything is expressed in terms of the eigenfunction $\Phi (0)$
of the ``initial'' Lax
operator $L(-1)$. The difference with the two-matrix case is only the explicit
form of the latter (recall $x \equiv \htt_1$) :
\be
L(-1) = e^{\g_{10} \htt_1} \( \sum_{s=0}^{m(1)} \a_s \g_{11}^{-s} D^s \)
e^{-\g_{10} \htt_1}
\lab{L-1-q-psi}
\ee
where the coefficients $\a_s ,\g_{10},\g_{11}$ have more complicated
dependence on $\{ \htt_s\}$ than in the two-matrix case.

Exactly as in the two-matrix case, we obtain the relation between the $n$-th
step DB eigenfunction $\Phi (n)$ and the orthogonal polynomial
normalization factor $h_n$ \rf{h-q-n} which generalizes \rf{h-Phi-n} :
\be
\Phi (n) \equiv e^{\int \hb_0 (n)} = h_n \, \g_{11}^n \,
\exp \lcurl \htt_1 \g_{10} + \vareps (\htt^{\pr}) \rcurl  \qquad ,\quad
\htt^{\pr} \equiv \( \htt_2 ,\ldots \htt_{m(q)}\)
\lab{h-Phi-n}
\ee
Substituting \rf{h-Phi-n} into \rf{ZN-h} and using the Wronskian formula
\rf{tau-W-n} we get:
\br
Z_N = \prod_{n=0}^{N-1} = \det {\Bigl\Vert}
\partderh{}{\htt_1}{i-1} \( L(-1)\)^{j-1} \Phi (0) {\Bigr\Vert}
e^{-N\( \htt_1 \g_{10} + \vareps (\htt^{\pr})\)} \g_{11}^{-\frac{N(N-1)}{2}}
\lab{ZN-det-q-0}  \\
= \det {\Bigl\Vert}
\partderh{}{\tit_1}{i-1}
\( e^{- \htt_1 \g_{10}} L(-1) e^{\htt_1 \g_{10}}\)^{j-1} h_0  {\Bigr\Vert}
\lab{ZN-det-q-1}
\er
where we absorbed the $\g_{11}$-factors via changing
$\partder{}{\htt_1} \to \partder{}{\tit_1}$ by the definition \rf{tatko-q-a},
{\sl i.e.}, $\g_{11}^{-1} \partder{}{\htt_1} = \partder{}{\tit_1}$.
Now, we find using \rf{L-1-q-psi} and \rf{h-0-constr} :
\be
\( e^{- \htt_1 \g_{10}} L(-1) e^{\htt_1 \g_{10}}\)^{j_1} h_0 =
\( \sum_{s=0}^{m(1)} \a_s \partderh{}{\tit_1}{s} \)^{j-1} h_0 =
\partderh{}{t_1}{j-1} h_0
\lab{L-1-j-h}
\ee
Substituting \rf{L-1-j-h} into \rf{ZN-det-q-1} yields the final result for the
multi-matrix model partition function:
\be
Z_N =\det {\Bigl\Vert}
\partderM{h_0}{i+j -2}{\tit_1}{i-1}{t_1}{j-1} {\Bigr\Vert}
\lab{ZN-det-q}
\ee
which is functionally the same as for the two-matrix model,
however, with a more complicated expression for $h_0$ \rf{h-q-n} :
\be
h_0 = \int_{\Gamma} \int_{\Gamma} d\l d\m \,
\exp \Bigl\{ \sum_{r=1}^{p_1} \l^r t_r \Bigr\}
\rho (\l ,\m ; \{ t^{\pr\pr}\}, \{ g\})
\Bigl\{ \sum_{s=1}^{p_q} \m^s \tit_s \Bigr\}
\lab{h-q-0}
\ee
Eq.\rf{ZN-det-q} was previously obtained (see refs.\ct{Moroz}) from a
different approach.
\lskip
{\bf Acknowledgements}\\
E.N. gratefully acknowledges support and
hospitality of the {\em Deutscher Akademischer Austauschdienst} and Prof.
K.Pohlmeyer at the University of Freiburg. S.P. acknowledges support from the
Ben-Gurion University, Beer-Sheva.


\small
\begin{thebibliography}{99}
\bibitem{Faddeev}
L.D. Faddeev and L.A.  Takhtajan, {\em ``Hamiltonian Methods in the Theory of
Solitons''}, Springer (1987)
\bi{QISM}
L. Faddeev, {\sl Integrable Models In 1+1 Dimensional Quantum Field Theory},
Les Houches Lectures, session XXXIX (1982),
J.-B. Zuber and R. Stora eds., Elsevier Sci. Publ. (1984); \\
L. Faddeev, in {\em Fields and Particles},
H. Mitter and W. Schweiger eds., Springer (1990); \\
P. Kulish and E. Sklyanin,
in {\sl Lect. Notes in Phys.} {\bf 151}, p.67, Springer (1982);  \\
L. Takhtajan, in {\em Introduction
to Quantum Group and Integrable Massive Models of Quantum Field Theory},
Mo-Lin Ge and Bao-Heng Zhao eds., World Sci. (1990)
\bi{Olsha}
M. Olshanetsky and A. Perelomov, \InvM{54}{1979}{261};
\TMP{45}{1980}{843}; \PR{71}{1981}{313}; {\sl ibid} {\bf 94} (1983); \\
{\sl ``B\"acklund transformations''}, Lecture notes in Mathematics,
Vol. 515, Eds. A. Dold and B. Eckmann, Springer (1976);  \\
{\sl ``Solitons''},
eds. R. Bullough and P. Caudrey, {\sl Topics in Current Physics},
Springer (1980)
\bi{Seiberg-Witten}
A. Gorsky, I. Krichever, A. Marshakov, A. Mironov and A. Morozov, in
{\sl hep-th/9505035}; \\
H. Itoyama and A. Morozov, in {\sl hep-th/9511126; hep-th/9512161}; \\
E. Martinec, in {\sl hep-th/9510204}; \\
E. Martinec and N. Warner, in {\sl hep-th/9511052}
\bi{W-inf}
V. Kac and D. Peterson, {\sl Proc. Natl. Acad. Sci. USA} {\bf 78} (1981)
3308;\\
I. Bakas, \CMP{134}{1990}{487}; \\
C. Pope, L Romans and X. Shen, \NPB{339}{1990}{191}; \\
A. Radul, \FAaIA{25}{1991}{33}; \\
A. Radul and I. Vaysburd, \PLB{274}{1992}{317}
\bi{Zam}
A. Zamolodchikov, \TMP{65}{1985}{1205}; \\
V. Fateev and S. Lukianov, \IJMPA{7}{1992}{853}; \\
P. Bouwknegt and K. Schoutens, \PR{223}{1993}{183}
\bibitem{W-appl}
(a) J. Ellis, N. Mavromatos and D. Nanopoulos, {\sl Phys. Lett.} {\bf 2647B}
(1991) 465; {\em ibid} {\bf 272B} (1991) 261; {\sl hep-th/9403133}  \\
(b) I. Klebanov and A. Polyakov, {\sl Mod. Phys. Lett.} {\bf A6} (1991) 3273;
E. Witten, {\sl Nucl. Phys.} {\bf B373} (1992) 187  \\
(c) Y. Watanabe, Annali Matem. Pura Appl. {\bf 136} (1984) 77; \\
A. Reiman and M. Semenov-Tyan-Shansky, J. Sov. Math. {\bf 31} (1985) 3399; \\
K. Yamagishi, Phys. Lett. {\bf 259B} (1991) 436 ; F. Yu and Y.-S. Wu,
{\em ibid} {\bf 263B} (1991) 220  \\
(d) Q. Han-Park, Phys. Lett. {\bf 236B} (1990) 429; {\em ibid} {\bf 238B} 208 ;
\IJMPA{7}{1992}{1415}; \\
K. Yamagishi and G. Chapline, Class. Quant. Grav. {\bf 8} (1991) 1 \\
(e) H. Ooguri and C. Vafa, \NPB{361}{1991}{469}; {\em ibid.} {\bf B367} (1991)
 83
\bibitem{Cappelli}
A. Cappelli, C, Trugenberger and G. Zemba, \NPB{396}{1993}{465};
\PRL{72}{1994}{1902}; \NPB{448 [FS]}{1995}{470};
{\sl hep-th/9502021; hep-th/9502050}; \\
D. Karabali, \NPB{428}{1994}{1994}; \\
B. Sakita, \PLB{315}{1993}124; \\
R. Ray and B. Sakita, {\sl Ann. Phys.} {\bf 230} (1994) 131
\bibitem{AKS}
B. Kostant, {\sl London Math.Soc.Lect.Notes}, Ser. {\bf 34} (1979) 287; \\
M. Adler, {\sl Inv. Math.} {\bf 50} (1979) 219; \\
W. Symes, {\sl Inv. Math.} {\bf 59} (1980) 13; \\
A.G. Reyman and M.A. Semenov-Tian-Shansky, {\sl Inv. Math.} {\bf 54}
(1979) 81; {\em ibid} {\bf 63} (1981) 423
\bi{Zakh}
S. Manakov, S. Novikov, L. Pitaevski and V. Zakharov, {\em ``Soliton Theory:
The Inverse Problem''}, Nauka, Moscow (1980)
\bi{Dickey}
L.A. Dickey, {\em ``Soliton Equations and Hamiltonian Systems''}, World
Scientific, Singapore (1991)
\bi{integr-matrix}
M. Douglas, \PLB{238}{1991}{176}; \\
E. Martinec, \CMP{138}{1991}{437}; \\
A. Gerasimov, Yu. Makeenko, A. Marshakov, A. Mironov, A. Morozov and A. Orlov,
\MPLA{6}{1991}{3079}; \\
R. Dijkgraaf, E. Verlinde and H. Verlinde, \NPB{348}{1991}{435}; \\
L. Alvarez-Gaum\'e, C. Gomez and J. Lacki,
{\sl Phys. Lett.} {\bf B253} (1991) 56
\bi{add-symm}
A. Orlov and E. Schulman, \LMP{12}{1986}{171}; \\
A. Orlov, in {\em Plasma Theory and Nonlinear and
Turbulent Processes in Physics\/} (World Scientific, Singapore, 1988); \\
L. Dickey, \CMP{167}{1995}{227} (also in {hep-th/9312015});
\MPLA{8}{1993}{1357} (also in {\sl hep-th/9210155})  \\
M. Adler, T. Shiota and P. van Moerbecke, \PLA{194}{1994}{33}; \\
L. Haine and E. Horozov, {\sl Bull. Sc. Math.}, 2-e serie, {\bf 117} (1993)
485
\bi{DB}
V. Matveev and M. Salle, {\em ``Darboux Transformations and Solitons''},
Springer-Verlag (1991); \\
A. Leznov, A. Shabat and R. Yamilov, \PLA{174}{1993}{397}; \\
W. Oevel, \PHSA{195}{1993}{533}; \\
L.-L. Chau, J.C. Shaw and H.C. Yen, \CMP{149}{1992}{263}
\bi{STS83}
M. Semenov-Tyan-Shansky, \FAaIA{17}{1983}{259}
\bi{W-h-inf}
I. Bakas and E. Kiritsis, {\sl Int. J. Mod. Phys.} {\bf A7} [Suppl. 1]
(1992) 55; \\
F. Yu and Y.-S. Wu, \NPB{373}{1992}{713}
\bi{chengs}
Y. Cheng, W. Strampp and B. Zhang, \CMP{168}{1995}{117}
\bi{avoda}
H. Aratyn, E. Nissimov and S. Pacheva, \PLA{201}{1995}{293}
(also in {\sl hep-th/9501018})
\bi{multi-b}
L. Bonora and C.S. Xiong, \PLB{317}{1993}{329};\\
H. Aratyn, E. Nissimov, S. Pacheva and I. Vaysburd,
{\sl Phys. Lett.} {\bf 294B} (1992) 167 (also in {\sl hep-th/9209006})
\bi{no2rabn1}
H. Aratyn, E. Nissimov and S. Pacheva,
{\sl Phys. Lett.} {\bf 331B} (1994) 82 (also in {\sl hep-th/9401058})
\bi{office}
H. Aratyn, E. Nissimov, S. Pacheva and A.H. Zimerman,
\IJMPA{10}{1995}{2537} (also in {\sl hep-th/9407117});
\bi{rio}
H. Aratyn, {\sl Lectures at the VIII J.A. Swieca Summer
School, 1995} (also in {\sl hep-th/9503211})
\bi{Yu}
F. Yu, \LMP{29}{1993}{175} (also in {\sl hep-th/9301053})
\bi{multikp}
H. Aratyn, E. Nissimov and S. Pacheva,
{\sl Phys. Lett.} {\bf 314B} (1993) 41 (also in {\sl hep-th/9306035})
\bi{Wronski}
E.L. Ince, {\sl Ordinary Differential Equations}, chap. V, London, 1926;\\
M.M. Crum, {\sl Quart. J. Math. Oxford} {\bf 6} (1955) 121;\\
M. Adler and J.Moser, \CMP{61}{1978}{1}
\bi{BX}
L. Bonora and C.S. Xiong, \IJMPA{8}{1993}{2973};
\NPB {405}{1993}{191} (also in {\sl hep-th/9212070});
\JMP{35}{1994}{5781} (also in {\sl hep-th/9311070});
``{\sl Correlation functions of two-matrix models}", in {\sl hep-th/9311089}
\bi{BX-2}
L. Bonora  and C.S. Xiong,
{\sl Phys. Lett.} {\bf B347} (1995) 41 (also in {\sl hep-th/9405004});
{\sl Nucl. Phys.} {\bf B434} (1995) 408 (also in {\sl hep-th/9407141});\\
L. Bonora, C. Constantinidis and E. Vinteler, in {\sl hep-th/9511172}
\bi{ortho-poly}
D.Bessis, C. Itzykson and J.-B. Zuber, {\sl Adv. Appl. Math.}{\bf 1} (1980)
109; \\
C. Itzykson and J.-B. Zuber, \JMP{21}{1980}{411}; \\
M. Mehta, \CMP{79}{1981}{327}; \\
S. Chadha, G. Mahoux and M. Mehta, \JPA{14}{1981}{579}
\bi{U-T}
K. Ueno and K. Takasaki, {\sl Adv. Stud. Pure Math.} {\bf 4} (1984) 1
\bi{oakpark}
H. Aratyn, E. Nissimov and S. Pacheva, preprint BGU-95/19/Dec-PH
(also in {\sl solv-int/9512nnn})
\bi{Moroz}
A. Morozov, {\sl Physics Uspekhi} {\bf 37} (1994) 1 (also in
{\sl hep-th/9303139}); in {\sl hep-th/9502091}

\end{thebibliography}
\end{document}